\documentstyle[preprint,prl,aps,epsfig]{revtex}
\def\jcp#1#2#3{J.~Chem.~Phys.~{\bf #1},\ #2\ (#3)}

\def\pra#1#2#3{Phys.~Rev.~A~{\bf #1},\ #2\ (#3)}
\def\prl#1#2#3{Phys.~Rev.~Lett.~{\bf #1},\ #2\ (#3)}

\def\adamo#1#2#3{Adv.~At.~Mol.~Opt.~Phys. {\bf #1},\ #2\ (#3)}
\newcommand{\beq}{\begin{equation}}
\newcommand{\eeq}{\end{equation}}
\begin{document}

\tightenlines

\flushbottom \draft
\title{
Breaking van der Waals molecules with magnetic fields
}
\author{R.V. Krems}
\address{ Harvard-MIT Center for Ultracold Atoms, Department of Physics, 
Harvard University and Institute for Theoretical Atomic, Molecular and Optical 
Physics, Harvard-Smithsonian Center for Astrophysics, Cambridge, MA 02138
\\
Electronic mail: rkrems@cfa.harvard.edu
 \\ (\today)
\\ \medskip}\author{\small\parbox{14.2cm}{\small\hspace*{3mm}
It is demonstrated that weakly bound van der Waals complexes 
can dissociate in a magnetic field through coupling between 
the Zeeman levels. The Zeeman predissociation process is shown 
to be efficient and it 
can be controlled by external magnetic fields. 
\\
\\[3pt]PACS numbers: 34.50.-s, 32.80.Pj, 32.60.+i, 34.20.6j }}
\maketitle
\maketitle \narrowtext

\vspace{1.cm}

Manipulating dynamics of molecules with external 
fields has long been a sought-after goal of experimental and 
theoretical research. 
Several groups have studied elastic and inelastic collisions 
of ultracold alkali atoms in the presence of Feshbach scattering resonances
\cite{e1,e2,e3,e4,e5,e6}. It was found that the probabilities of 
elastic collisions and inelastic 
energy transfer undergo dramatic changes near Feshbach 
resonances. The Feshbach resonances can be tuned by bias 
magnetic fields 
so that atom-atom ultracold collisions can be controlled. 
Ultracold molecules have been created by linking ultracold atoms 
with magnetic fields \cite{e7}. 
The creation of molecules 
at subKelvin temperatures has opened up possibilities 
for controlling molecular interactions with electric fields 
\cite{john_bohn}. External fields break the isotropy 
of space and may induce couplings between 
electronic states otherwise uncoupled \cite{roman1}. 
Forbidden electronic transitions may thus become 
allowed in an external field and the probabilities of 
the electronic transitions can be controlled by 
the strength of the external field \cite{roman1}. 

In this work we study predissociation 
of weakly bound van der Waals complexes containing atoms 
with non-zero electronic orbital angular 
momentum in a magnetic field
when transitions to 
lower magnetic levels
release enough energy to 
break the van der Waals bond. 
The predissociation is 
due to coupling 
between different 
Zeeman energy levels of the complex. 
We show that the Zeeman predissociation 
dynamics can be efficient and the molecules, 
while stable at zero magnetic field, 
quickly decompose in a magnetic field. 

Figure 1 illustrates the idea of the Zeeman 
predissociation. Consider a van der Waals complex of an 
open-shell atom (A) with non-zero electronic orbital
angular momentum and a closed-shell atom or molecule (B) 
in a magnetic field. 
The AB complex can be described by a set of coupled 
potential energy 
curves correlating with different Zeeman energy levels 
of atom A. At zero magnetic field, the Zeeman 
energy levels are degenerate and the complex 
is stable.   
An external magnetic field
splits the Zeeman energy levels and bound states of 
the complex 
may be higher in energy than 
some of the Zeeman levels of the separated A and B. 
The AB molecule can then 
dissociate through transitions to these Zeeman levels. 
The predissociation lifetime 
depends on the strength of the coupling between 
the Zeeman levels. 

If atom A is in an electronic $S$-state, the A-B interaction is 
independent of spin and  
the Zeeman energy levels
are not coupled but they are if atom A is  
 in a state with non-zero 
orbital angular momentum. If B is a closed-shell atom, the A-B 
interaction potential can be expanded in a Legendre series \cite{alex}

\begin{equation}
V_{\rm AB}({\bf R}, \hat{r}) = \sum_{\lambda} \frac{4\pi}{2\lambda + 1}
V_{\lambda}(R) \sum_{m_{\lambda}}
Y^*_{\lambda m_{\lambda}}(\hat{R})Y_{\lambda m_{\lambda}}(\hat{r}),
\label{pot}
\end{equation}

\noindent
where $R$ is the interatomic distance, $\hat{R}$ is the unit vector with 
the direction of ${\bf R}$, 
${\hat r}$ denotes collectively 
the position vectors of the open-shell 
electrons and the expansion coefficients 
$V_{\lambda}(R)$ can be expressed in terms of the non-relativistic 
interaction potentials of the AB molecule \cite{aquil}.
The vector sum of the electronic orbital
${\bf L}$ and spin ${\bf S}$ angular momenta gives the total 
electronic angular momentum ${\bf J}_{\rm a}$ of atom A. 
The asymptotic $R = \infty$ states of the AB complex in a magnetic field 
are characterized by the rotational angular momentum $l$ of AB and
the projections $m_l$ and $M$
of ${\bf l}$ and ${\bf J}_{\rm a}$ 
on the magnetic field axis, respectively \cite{roman1}. 
The magnetic field couples the states of different total angular momentum 
$J_{\rm a}$, but not the $M$-states, so that the Zeeman transitions 
are induced by the couplings between states $|J_{\rm a} M l m_l {\rangle}$
and $|J_{\rm a}' M' l' m_l' {\rangle}$. The integrals over 
the wavefunctions $|J_{\rm a} M l m_l {\rangle}$
with the interaction potential (\ref{pot}) can be evaluated analytically 
to yield \cite{roman1}

\begin{eqnarray}
\nonumber
{\langle} J_{\rm a}(LS)M l m_l |V_{\rm AB} | J_{\rm a}'(LS)M'l'm_l'{\rangle} = \sum_{\lambda} 
V_{\lambda} \sum_{m_{\lambda}} (-1)^{S + J_{\rm a} + J_{\rm a}' + \lambda + m_{\lambda} - m_l - M}
\\
\nonumber
 [(2L+1)(2L+1)(2J_{\rm a}+1)(2J_{\rm a}'+1)(2l+1)(2l'+1)]^{1/2} {\times}
\\
 \left \{ \begin{array}{c c c} 
                                                 L  & J_{\rm a}   & S         \\
                                                 J_{\rm a}' & L   & \lambda   \\
                              \end{array} \right \}
 \left ( \begin{array}{c c c} 
                                                  J_{\rm a} & \lambda       & J_{\rm a}'         \\
                                                 -M & m_{\lambda}  & M'   \\
                              \end{array} \right )
 \left ( \begin{array}{c c c} 
                                                  l & \lambda       & l'         \\
                                                 -m_l & -m_{\lambda}  & m_l'   \\
                              \end{array} \right )
 \left ( \begin{array}{c c c} 
                                                  L & \lambda       & L         \\
                                                  0 & 0  & 0   \\
                              \end{array} \right )
 \left ( \begin{array}{c c c} 
                                                  l & \lambda       & l'         \\
                                                  0 & 0 & 0  \\
                              \end{array} \right ),
\label{vesmat}
\end{eqnarray}

\noindent
where symbols in parentheses and curly braces are $3j$- and $6j$-symbols. 
Eq. (\ref{vesmat}) establishes that the Zeeman energy levels are coupled 
by the anisotropic part of the interaction potential (\ref{pot}). 
Because the interaction of atoms with non-zero $L$ depends strongly
on the relative orientation of $\hat{R}$ and ${\hat r}$, 
the Zeeman couplings are significant in the van der Waals complexes 
with such atoms.

  To form an idea of lifetimes for the Zeeman predissociation
in complexes with open-shell atoms, we analyzed
the predissociation of the He-O($^3P$) molecule in a magnetic field. 
The ground state of oxygen is $^3P_2$. It splits 
into five Zeeman levels with energies approximately 
equal to $3 B \mu_B M /2$, where $\mu_B \approx 0.47$ cm$^{-1}$/Tesla. 
This suggests that the HeO molecules bound by $11.2$ cm$^{-1}$ 
or less may undergo the Zeeman predissociation at $B=4$ Tesla. 
The ground state of the He-O($^3P$) complex is characterized by two
non-relativistic potentials of $\Sigma$ and 
$\Pi$ symmetry \cite{roman2}.
Although the well depths of the potentials are only 
$9.4$ and $20.9$ cm$^{-1}$, respectively, they support several bound states.
To identify the predissociating states, we computed the $S$-matrix
eigenphase sum for He - O($^3P$) collisions 
on a dense grid of energies below the $M = + 2$ threshold. 
The theory of the scattering calculations has been described 
in our preceding paper \cite{roman1}. 
The eigenphase sum is obtained by summing 
the inverse tangents of the eigenvalues of the $K$-matrix \cite{jeremy2}.
The eigenphase sum rises with energy by $\pi$ 
near an isolated resonance \cite{hazi}. 

  Figure 2 shows the eigenphase sum for He-O($^3P$) as a function of 
energy and external magnetic field.
When oxygen is in the $^3P_2$ state, there 
are no couplings between the states of odd and even 
partial waves $l$ ({\it cf.} Eq. \ref{vesmat}).
Five Zeeman levels corresponding to $J_{\rm a}=2$ and 
three states of the orbital angular momentum, $l=0,2$ and $4$,
are included in these calculations with the fixed 
total angular momentum projection $M + m_l = 2$.
The spin-orbit excited states of oxygen are separated from the 
$J_{\rm a}=2$ state by $158.1$ and $226.6$ cm$^{-1}$, 
so they were neglected.
The number of resonances becomes larger 
and the resonances broaden with increasing magnetic field. The lifetimes of the metastable states 
can be computed from the energy derivative of the eigenphase sum 
 at the positions of the resonances \cite{jeremy2}. To make sure that our calculations 
are correct, we computed the derivatives both by the finite difference method 
and analytically from piecewise Chebyshev approximations of the eigenphase sum. 
The positions and lifetimes of the resonances of Fig. 2 are summarized in Table 1. 
The lifetimes decrease by almost two orders of magnitude as the magnetic field 
increases from $0.1$ to $3$ Tesla.

The rise of the eigenphase sum in Fig. 2 is due to the predissociating states 
or shape resonances in open channels. To verify that we do observe 
the Zeeman predissociation, we computed the bound states of 
the O($^3P_2$)-He complex as a function of the magnetic field. 
To do that, we diagonalized the Hamiltonian of the 
O($^3P_2$)-He complex at zero magnetic field using a Hund's case (e)
representation for the atomic wave functions \cite{jeremy3}
and a Fourier basis discrete variable representation (DVR) of Colbert and Miller \cite{miller}
for the radial part of the wave function.  
The bound states are characterized 
by the value of total angular momentum ${\bf J} = {\bf J}_{\rm a} + {\bf l}$, 
parity $\epsilon = (-1)^{L + l}$ and a quantum number $s$ describing 
the stretching motion of the nuclei (see Table 2).  
To confirm the accuracy of our results, we repeated the 
calculation in the Hund's case (c) representation of the 
molecular functions.  The total Hamiltonian was then diagonalized 
in the fully uncoupled representation (\ref{vesmat})
as a function of the magnetic field strength. 
This procedure yields converged results for the bound energy levels and 
the predissociating states with lifetime $> 10^{-9}$ sec, which was verified by 
the close coupling calculations. As the lifetime of the predissociating states becomes
smaller than $10^{-9}$ sec, the DVR calculations converge very slowly and the close coupling
calculation was used to obtain the magnetic field dependence of the predissociating states 
at high magnetic fields. The close coupling solution at high magnetic fields was matched 
with the DVR results at low fields to provide the magnetic field dependence of the energy 
levels in the full interval of magnetic fields considered. This allowed us to identify three 
resonances in Fig. 1 and Table 1 with the quantum numbers $J_{\rm a}$ and $s$. 
The fourth resonance is a shape resonance.

 When the strength of the magnetic field increases, more Zeeman levels become energetically 
accessible and the dissociation can occur through transitions to more states. 
The energy gap between the initial and final energy levels is larger in higher 
magnetic fields and the wave packet escapes faster over the centrifugal barriers 
in the outgoing channels. At the same time, more bound states become open 
for predissociation in higher magnetic fields (see Fig. 1). 
Thus the number of metastable states and the 
dissociation efficiency increase with 
increasing field strength.

The lifetimes of the bound states decaying 
through the Zeeman predissociation are small (see Table 1) 
indicating that the Zeeman 
levels in He-O($^3P$) are strongly coupled. 
Yet, it should be expected that the degree of anisotropy
is smaller in the HeO complex than in most other van der Waals 
complexes containing atoms with non-zero $L$. 
The strength and anisotropy of interaction 
of oxygen with other rare gas atoms are larger \cite{roman2}.
The degree of 
anisotropy is also larger in complexes containing 
D-state atoms \cite{note}.

If A is an open-shell atom and B is a $\Sigma$-state diatomic molecule, 
the angular expansion of the A-B 
interaction potential is analogous to that of 
two diatomic molecules \cite{jeremy1}. 
It can be written in the form \cite{ad}

\begin{eqnarray}
\nonumber
V_{\rm AB}({\bf R}, {\hat r}_{\rm A}, {\bf r}_{\rm B}) = 
(4 \pi)^{3/2}
\sum_{\lambda_{\rm A} \lambda_{\rm B} \lambda} V_{\lambda_{\rm A} \lambda_{\rm B} \lambda} 
(R, r_{\rm B}) \hspace{1.cm}
\\
{\times} \sum_{ m_{\lambda_{\rm A}} m_{\lambda_{\rm B}} m_{\lambda}} 
  \left ( \begin{array}{c c c} 
                                    \lambda_{\rm A}      &   \lambda_{\rm B}      &  \lambda   \\
                                    m_{\lambda_{\rm A}}  &   m_{\lambda_{\rm B}}  &  m_{\lambda}   \\
                              \end{array} \right )
Y_{\lambda_{\rm A} m_{\lambda_{\rm A}}}(\hat{r}_{\rm A})
Y_{\lambda_{\rm B} m_{\lambda_{\rm B}}}(\hat{r}_{\rm B})
Y_{\lambda m_{\lambda}}(\hat{R}),
\label{pottwo2sig}
\end{eqnarray}

\noindent
where ${\hat r}_{\rm A}$ is the same as ${\hat r}$ in Eq. (\ref{pot}) and 
$\hat{r}_{\rm B}$ describes the orientation of molecule B in the space-fixed 
coordinate frame. The Zeeman energy levels are coupled by the matrix elements
between states $|NM_NJ_{\rm a} M l m_l{\rangle}$ and $|N'M_N'J_{\rm a}' M' l' m_l'{\rangle}$, 
where $N$ is the rotational angular momentum quantum number and $M_N$ is 
the projection of ${\bf N}$. The matrix 
elements of the interaction potential can be evaluated using the 
Wigner-Eckart theorem as follows

\begin{eqnarray}
\nonumber
{\langle} NM_NJ_{\rm a} M l m_l | V_{\rm AB} | N'M_N'J_{\rm a}'M' l' m_l' {\rangle} = 
\sum_{\lambda_{\rm A} \lambda_{\rm B} \lambda} V_{\lambda_{\rm A} \lambda_{\rm B} \lambda} 
\hspace{1.cm}
\\
\nonumber
{\times} \sum_{ m_{\lambda_{\rm A}} m_{\lambda_{\rm B}} m_{\lambda}} 
  \left ( \begin{array}{c c c} 
                                    \lambda_{\rm A}      &   \lambda_{\rm B}      &  \lambda   \\
                                    m_{\lambda_{\rm A}}  &   m_{\lambda_{\rm B}}  &  m_{\lambda}   \\
                              \end{array} \right )
(-1)^{S + J_{\rm a} + J_{\rm a}' + \lambda_{\rm A} - m_l - M - M_N}
\\
\nonumber
[(2L+1)(2L+1)(2J_{\rm a}+1)(2J_{\rm a}'+1)(2l+1)(2l'+1)(2N + 1)(2N'+1)]^{1/2}
\\
\nonumber
[(2\lambda_{\rm A}+1) (2 \lambda_{\rm B} + 1)(2\lambda+1)]^{1/2}
 \left \{ \begin{array}{c c c} 
                                                  L & J_{\rm a}   & S         \\
                                                 J_{\rm a}' & L   & \lambda_{\rm A}   \\
                              \end{array} \right \}
 \left ( \begin{array}{c c c} 
                                                  J_{\rm a} & \lambda_{\rm A}       & J_{\rm a}'         \\
                                                 -M & m_{\lambda_{\rm A}}  & M'   \\
                              \end{array} \right )
 \left ( \begin{array}{c c c} 
                                                  l & \lambda       & l'         \\
                                               -m_l & m_{\lambda}  & m_l'   \\
                              \end{array} \right )
\\
{\times}
 \left ( \begin{array}{c c c} 
                                                  L & \lambda_{\rm A}       & L         \\
                                                  0 & 0  & 0   \\
                              \end{array} \right )
 \left ( \begin{array}{c c c} 
                                                  l & \lambda       & l'         \\
                                                  0 & 0 & 0  \\
                              \end{array} \right )
 \left ( \begin{array}{c c c} 
                                                  N & \lambda_{\rm B}       & N'         \\
                                                  -M_N & m_{\lambda_{\rm B}} & M_N'  \\
                              \end{array} \right )
 \left ( \begin{array}{c c c} 
                                                  N & \lambda_{\rm B}       & N'         \\
                                                  0 & 0 & 0  \\
                              \end{array} \right )
\label{zeemanMol}
\end{eqnarray}

\noindent
Eq. (\ref{zeemanMol}) shows that in addition to the anisotropy of interaction due to relative 
rotation of vectors ${\hat r}_{\rm A}$ and ${\hat R}$, the Zeeman transitions 
can be induced by the anisotropy of interaction arising from relative rotation 
of vectors ${\hat r}_{\rm A}$ and ${\hat r}_{\rm B}$. 
While the Zeeman transitions in atom-atom systems described by Eq. (\ref{pot})
must always be accompanied by changes of the orbital angular momentum projection 
$m_l$, the Zeeman transitions in atom-molecule collisions may occur without any 
change of $m_l$. The orientation of the electronic momentum ${\bf J}_{\rm a}$ may be 
transferred to the rotational angular momentum ${\bf N}$ of the molecule. 
The interaction between the Zeeman levels 
will therefore be generally stronger in complexes 
of open-shell atoms with molecules.

In summary, we have demonstrated that weakly bound van der Waals complexes 
containing atoms with non-zero electronic orbital angular momentum 
can dissociate in a magnetic field through coupling between the Zeeman 
energy levels. The lifetimes for the Zeeman predissociation are small
at laboratory moderate magnetic fields.  
They are similar to the lifetimes for rotational predissociation 
of triatomic van der Waals complexes such as Ar-HCl \cite{jeremy2}. 
Unlike ro-vibrational or electronic predissociation, the Zeeman 
predissociation can be controlled by an external magnetic field.

I would like to thank Alex Dalgarno for encouragement and 
useful comments on the manuscript 
and Sture Nordholm for hospitality during a visit to 
G\"{o}teborg University where this work was initiated. 
This work was supported by the Harvard-MIT 
Center for Ultracold Atoms and the Institute for Theoretical Atomic, 
Molecular and Optical Physics at the Harvard-Smithsonian Center 
for Astrophysics.

\begin{table}
\begin{center}
\caption{Positions ($E_{\rm r}$ in cm$^{-1}$) and lifetimes ($\tau_{\rm r}$ in sec) 
of the resonances of Fig. 2.
The energy is referred to the ($J_{\rm a}, M=+2$) dissociation limit. 
}

\begin{tabular}{c c c c c}
 $B$, Tesla  &    $J$   &  $s$  & $E_{\rm r}$  & $\tau_{\rm r}$   \\  \hline
 $0.1$       &          &       & $-0.003$     &    $1.14{\times}10^{-9}$ \\ 
\\
 $0.3$       &    2     &    1  & $-0.298$     &    $3.63{\times}10^{-9}$ \\
             &          &       & $-0.131$     &    $3.41{\times}10^{-10}$ \\
\\
 $1.0$       &    3     &    0  & $-1.914$     &    $1.91{\times}10^{-8}$ \\
             &    2     &    1  & $-0.648$     &    $3.04{\times}10^{-9}$ \\
             &          &       & $-0.348$     &    $5.31{\times}10^{-11}$ \\
\\
 $2.0$       &    2     &    0  & $-4.023$     &    $4.33{\times}10^{-9}$ \\
             &    3     &    0  & $-2.814$     &    $8.54{\times}10^{-10}$ \\
             &    2     &    1  & $-1.248$     &    $1.84{\times}10^{-10}$ \\
             &          &       & $-0.403$     &    $4.00{\times}10^{-11}$ \\
\\
 $3.0$       &    2     &    0  & $-5.180$     &    $1.11{\times}10^{-10}$ \\
             &    3     &    0  & $-3.344$     &    $4.49{\times}10^{-11}$ \\
             &    2     &    1  & $-1.781$     &    $4.10{\times}10^{-11}$ \\
             &          &       & $-0.422$     &    $3.00{\times}10^{-11}$ \\
\\
\end{tabular}
\end{center}
\end{table}

\begin{table}
\begin{center}
\caption{
Bound energy levels ($E_r$ in cm$^{-1}$) of the O($^3P_{2}$)-He complex at zero magnetic field.
The energy is referred to the dissociation limit.
}

\begin{tabular}{c c c c}

$J$  &  $s$   &  $\epsilon$  & $E_r$  \\ \hline

$1$  &  $0$   &  $+1$        & $-0.701$  \\
\\
$2$  &  $0$   &  $-1$        & $-3.170$  \\
$2$  &  $0$   &  $+1$        & $-3.007$  \\
$2$  &  $1$   &  $-1$        & $-0.176$  \\
\\
$3$  &  $0$   &  $+1$        & $-1.622$  \\ 
$3$  &  $0$   &  $-1$        & $-0.947$  \\
\hline
 \end{tabular}
\end{center}
\end{table}

\begin{figure}
\begin{center}
\epsfig{file=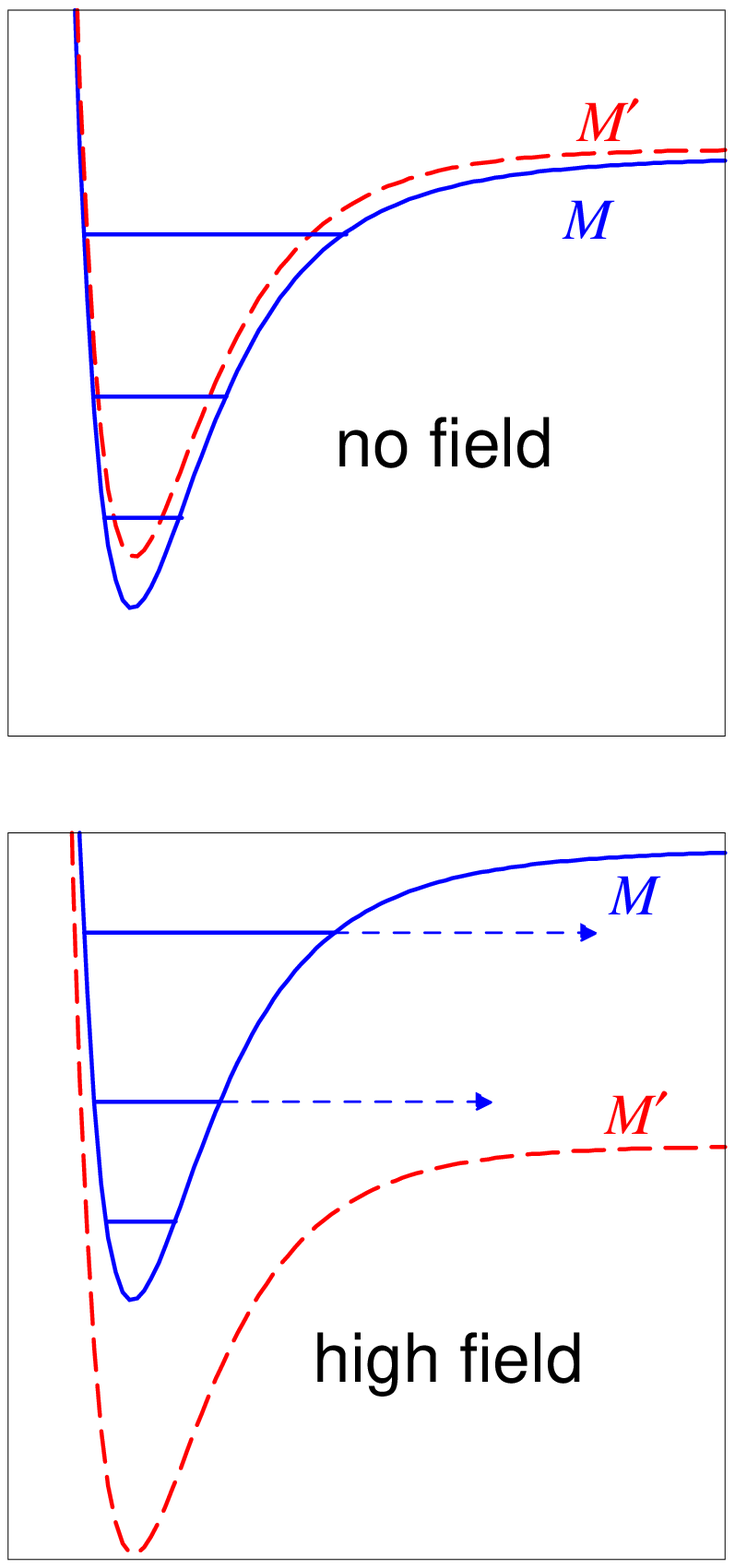,width=7.cm,angle=0}
\caption{ 
Zeeman predissociation
} \label{f1}
\end{center}
\end{figure}

\begin{figure}
\begin{center}
\epsfig{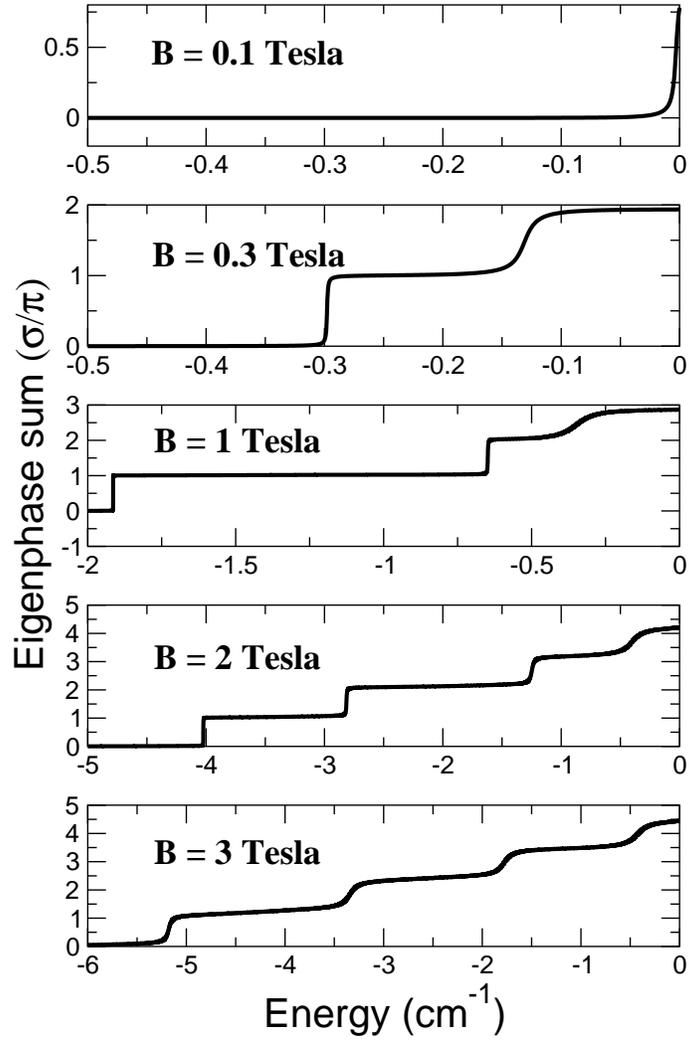}
\vspace{1.2cm}
\caption{ 
$S$-matrix eigenphase sum of He-O($^3P$) in a magnetic field.
The energy is referred to the $M=+2$ threshold.
} \label{f3}
\end{center}
\end{figure}


\begin{thebibliography}{99}
\bibitem{note}
This discussion excludes transition metal 
atoms. The interaction anisotropy
in complexes with transition metal atoms 
may be suppressed due to the peculiar character 
of their electronic structure.
\end{thebibliography}
\end{document}